SUPER-RESOLUTION MICROSCOPY

Correspondence and requests for materials should be addressed to P.X. (xipeng@pku.edu.cn)

# Fast Super-Resolution Imaging with Ultra-High Labeling Density Achieved by Joint Tagging Super-Resolution Optical Fluctuation Imaging

Zhiping Zeng[1][†], Xuanze Chen[1][†], Hening Wang[1], Ning Huang[2], Chunyan Shan[2], Hao Zhang[1], Junlin Teng[2] & Peng Xi[1][*]

[1]Department of Biomedical Engineering, College of Engineering, Peking University, Beijing, China

[2]College of Life Sciences, Peking University, Beijing, China

**Previous stochastic localization-based super-resolution techniques are largely limited by the labeling density and the fidelity to the morphology of specimen. We report on an optical super-resolution imaging scheme implementing joint tagging using multiple fluorescent blinking dyes associated with super-resolution optical fluctuation imaging (JT-SOFI), achieving ultra-high labeling density super-resolution imaging. To demonstrate the feasibility of JT-SOFI, quantum dots with different emission spectra were jointly labeled to the tubulin in COS7 cells, creating ultra-high density labeling. After analyzing and combining the fluorescence intermittency images emanating from spectrally resolved quantum dots, the microtubule networks are capable of being investigated with high fidelity and remarkably enhanced contrast at sub-diffraction resolution. The spectral separation also significantly decreased the frame number required for SOFI, enabling fast super-resolution microscopy through simultaneous data acquisition. As the joint-tagging scheme can decrease the labeling density in each spectral channel, we can faithfully reflect the continuous microtubule structure with high resolution through collection of only 100 frames per channel. The improved continuity of the microtubule structure is quantitatively validated with image skeletonization, thus demonstrating the advantage of JT-SOFI over other localization-based super-resolution methods.**

Optical fluorescence microscopy has been routinely applied to explore a vast variety of biological phenomena[1]. When it comes to unveil the inner world of cells, conventional optical microscopy has encountered exceptional challenges in discerning fine structures inside a cell due to the resolution barrier bestowed by optical diffraction. In recent years, quantum dots (QDs) exhibit great potential for fluorescence imaging in life sciences[2-4]. This can be attributed to the remarkable optical properties of QDs, e.g., higher fluorescence brightness, superior photostability, blue shifted absorption spectra and narrow fluorescence emission spectra. More importantly, fluorescence intermittency (i.e., blinking) is a significant characteristic of QDs[5-7].

In the past decade, various super-resolution techniques aiming at breaking the diffraction barrier have sprung up[8]. Based on their mechanism of surpassing the diffraction limit, super-resolution microscopy techniques can be categorized into two categories: 1) targeted modulation, such as stimulated emission depletion microscopy (STED)[9-11], saturated structured illumination microscopy (SSIM)[12], and 2) stochastic blinking/fluctuation modulation, such as photo-activated localization microscopy (PALM) [13] and stochastic optical reconstruction microscopy (STORM)[14], Super-resolution Optical Fluctuation Imaging (SOFI) [15-17], etc. While the single-molecule localization-based super-resolution techniques are able to achieve remarkably high resolution, its applicability for live cell imaging is significantly confined by the requirement that no close-by emitters can be switched on simultaneously. This has largely limited the labeling density of the single-molecule localization techniques[18, 19].

SOFI is a technique developed to take the advantage of the blinking mechanism to achieve background-free, contrast-enhanced fast super-resolution imaging[15, 20]. As it is based on the temporal and spatial cross-correlation analyses of fluorescence fluctuation, the blinking nature of the QDs can be strategically used. Comparing with the pure localization-based techniques, SOFI allows much higher labeling density owing to the robust correlation analysis for separating closely spaced emitters with blinking signals[17]. Labeling density is of great importance for ensuring the structural integrity presented by optical fluorescence microscopy[21]. However, with the increase of the labeling density, the high-order cumulants of SOFI algorithm tend to induce artifacts, degrading the image quality.

Herein, we propose a method that can enable ultra-high labeling density super-resolution imaging, meanwhile retains the continuities and integrity of the targets being investigated without compromising the spatial resolution enhancement, through spectral multiplexing[22-26] and SOFI imaging (Joint Tagging SOFI, i.e., JT-SOFI). In our experiment, multiple types of quantum dots with their fluorescence spectra well separated were jointly immuno-stained to the same cellular structure (microtubules) in COS7 cells. Under such circumstances, the labeling density of single color QDs is relatively low which facilitates the accurate separation of single QDs using SOFI algorithm, at relatively low frame numbers. Yet, the overall labeling density is m-fold increased through the application of m types of QDs, thereby enabling JT-SOFI nanometric imaging with ultra-high labeling density. Owing to the blue-enhanced absorption spectra and narrow fluorescence emission spectra of QDs, the QDs can be excited simultaneously with the same excitation source, with excessively high spectral encoding capability[27]. By combining the multiple spectral channels, super-resolution images with well-preserved integrity and continuities can be reconstructed, which are capable of revealing subdiffraction-sized structures inside the biological cells in a more genuine perspective at high spatiotemporal resolution.

**Results**

**Illustration of joint tagging protocol by multiple types of quantum dots.**

We introduce quantum dots joint tagging protocol that enables high-order SOFI processing (more resolution improvement can be obtained), while retaining the continuous biological structures without compromising the spatial resolution. As illustrated in Fig. 1a, multiple types of quantum dots are jointly tagged to the microtubule networks. As the QDs exhibit notably narrow fluorescence emission spectra, it can be well distinguished spectrally for simultaneous imaging. Fig. 1b shows the schematic comparison of single and joint tagging under ultra-high labeling density situations. When the microtubule network is tagged with one single type of QDs with excessively high labeling density, too many QDs are bound to blink simultaneously, resulting in redundant overlapping events, and the decrease in visibility for high-order SOFI. However, when the microtubule network is jointly tagged with multiple types of QDs, the labeling density for each single type of QDs is relatively low which enables high-order SOFI processing with less frame numbers for acquiring higher spatiotemporal resolution improvement; yet, the overall labeling density is m-fold higher, enabling us to produce a composite image with improved resolution enhancement while retaining the continuities, fidelity and integrity of the targets being investigated.

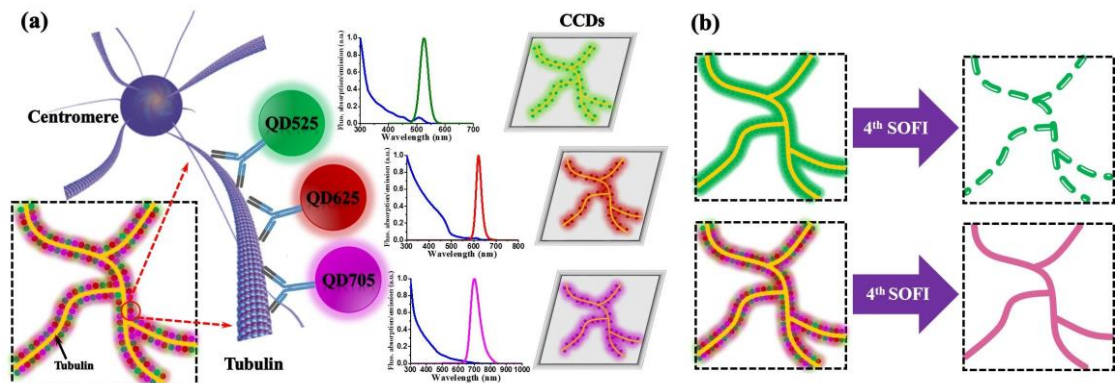

**Figure 1 Diagram of JT-SOFI.** (a) Illustration of joint tagging protocol by multiple types of quantum dots. (b) Schematic comparison of single and joint tagging under ultra-high labeling densities after high-order SOFI processing. Single tagging can be achieved by labeling any single type of QDs with ultra-high densities. Joint tagging can be achieved by jointly labeling QD525, QD625 and QD705 with relatively low densities for each channel.

**Simulation of blinking quantum dots with different labeling densities.** We simulated an image sequence of stochastically blinking emitters distributed on two closely spaced parallel lines in Fig. 2. The wavelength used in Figs. 2a-2g is 705 nm. The corresponding FWHM of Gaussian PSF is 250 nm. The wavelengths used in Figs. 2i, 2j and 2k are 525 nm, 625 nm and 705 nm, respectively. And the corresponding FWHMs of Gaussian PSFs in Figs. 2i, 2j and 2k are 190 nm, 225 nm and 250 nm, respectively. Figs. 2m-2o are merged images from Figs. 2i-2k. Pixel size of all images in Fig. 2 is 20 nm.

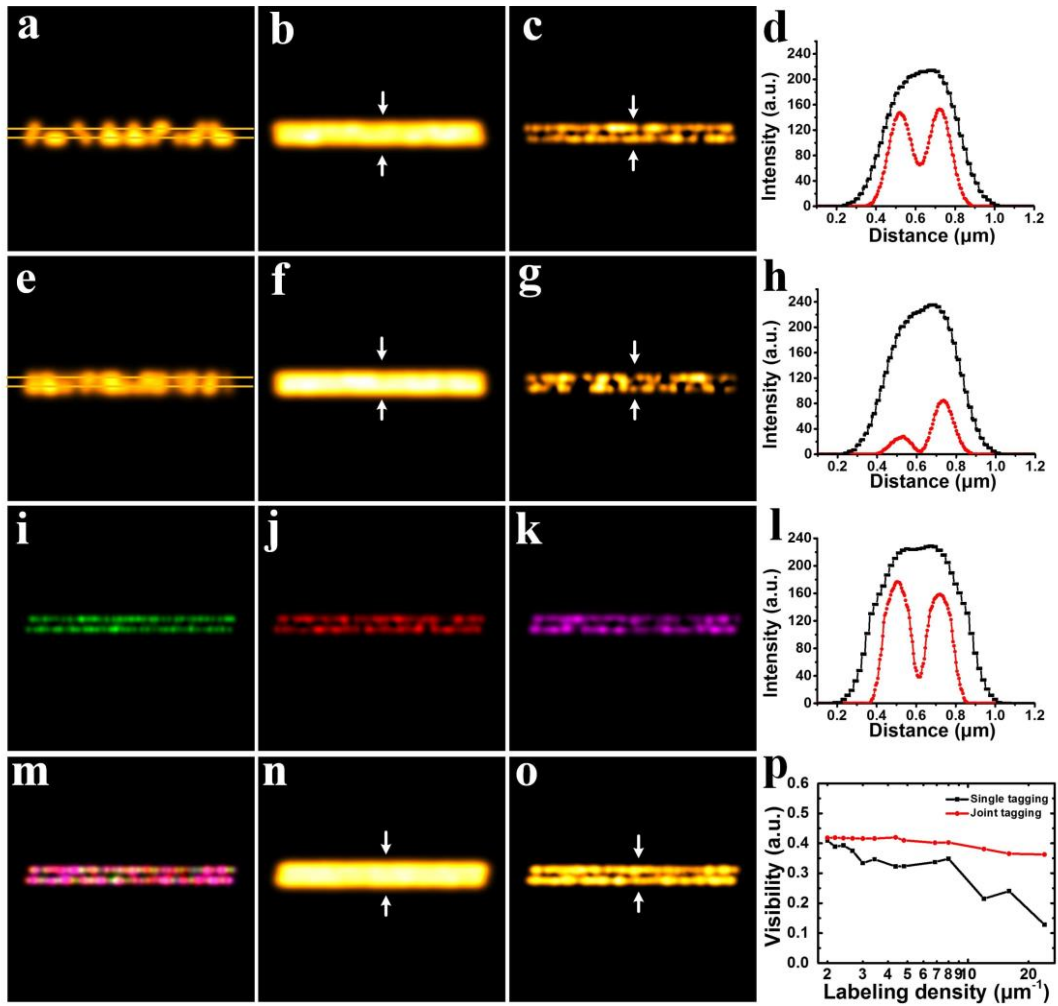

**Figure 2 Comparison of SOFI and JT-SOFI with simulation.** (a) A frame of simulation showing the random blinking characteristic of quantum dots distributed on two closely spaced parallel lines. (b) Average image of superimposing 100 independent blinking images in (a). (c) 3rd order SOFI super-resolution image of (a). (d) Cross-sections indicated by white arrows in (b) and (c). (e) A frame of simulation showing the blinking emitters at high labeling density (24 μm$^{-1}$). (f) Average image of superimposing 100 independent blinking images in (e). (g) 3rd order SOFI super-resolution image of (e). (h) Cross-sections indicated by white arrows in (f) and (g). (i)-(k) 3rd order SOFI super-resolution images of the green, red and magenta channels, respectively. (l) Cross-sections indicated by white arrows in (n) and (o). (m) Merged image of (i), (j) and (k). (n) Average image of joint tagging images. (o) 3rd order SOFI image of joint tagging images. (p) Visibility versus labeling density of single tagging and joint tagging schemes. The labeling densities in (a), (b), (c), (i), (j) and (k) are 8 μm$^{-1}$. The labeling densities in (e), (f), (g), (m), (n) and (o) are 24 μm$^{-1}$. Image size: 4.8 μm×4.8 μm.

As shown in Fig. 2, 100 frames were simulated for SOFI reconstruction. Due to the optical diffraction limit, the average result (Fig. 2b) produces an inauthentic image in which the blinking emitters seem to distribute on one single line. In Fig. 2c, after SOFI super-resolution reconstruction, it can be seen that the blinking emitters are distributed on two closely spaced parallel lines. However, under this insufficient labeling density situation, heterogeneities and discontinuities were undesirably introduced into the reconstructed image. In Fig. 2d, significant resolution improvement of 3rd order SOFI result over average counterpart can be observed. Subsequently, in Fig. 2e, the density of the blinking emitters is tripled. This setting is intended to simulate the situation where excessively high labeling density is applied. Fig. 2f shows the average image of superimposing 100 frames in Fig. 2e. Owing to the excessively densely packed emitters distributed on the closely spaced lines, the 3rd order SOFI result generated disastrous heterogeneities and artifacts shown in Fig. 2g. Interestingly, the SOFI process of the high labeling density gives no better resolution improvement, but increased heterogeneities and discontinuities over that in Fig. 2c. The above simulation has demonstrated that under the condition of ultra-high labeling density, high-order (typically over 3rd order) SOFI algorithm fails to significantly improve the spatial resolution.

In the subsequent simulations illustrated in Fig. 2i to 2o, the blinking emitters with different colors were predefined at different sets of coordinates along two parallel lines, simulating multi-color joint labeling scheme. As the sizes of PSFs vary in different wavelength, their sizes were set according to the corresponding spectrum. Each of the single color emitters in each particular spectral channel is loosely packed (8 $\mu m^{-1}$). However, when it comes to the overall density encompassing all the spectral channels, the multiple types of emitters are densely packed (24 $\mu m^{-1}$). As the density of each type of emitters is relatively low, SOFI processing produces a resolution-enhanced sub-image containing a fraction of the information (Figs. 2i, 2j and 2k). After combining all the SOFI-processed sub-images into one composite image (Fig. 2m), we are able to obtain a super-resolution image with high imaging fidelity and well retained continuities of the targeted structures (Fig. 2o).

In Fig. 2p, we simulated the visibility of SOFI reconstructed images versus labeling densities. The labeling density ranges from 2 to 24 µm$^{-1}$. Typically, if the labeling density exceeds 15 µm$^{-1}$, PALM/STORM algorithms completely fail to localize individual spots (the visibility falls to zero)[17]. As can be seen in Fig. 2p, for single tagging scheme, the visibility drops to 0.2 when the labeling density is over 15 µm$^{-1}$. Whereas, for joint tagging scheme, the degradation of visibility appears much less significant. The visibility is still above 0.35 even when the labeling density is over 24 µm$^{-1}$. Specifically, under high labeling density regime, joint tagging scheme exhibits extraordinarily better maintaining of the visibility over that in single tagging scheme. This simulation reveals the superior capability of joint tagging scheme for producing better SOFI images with higher visibilities. Therefore, JT-SOFI enables higher labeling density in biological applications, achieving better representation of the original biological fine structures.

**Imaging microtubule networks with quantum dots using multi-color joint tagging protocol.** A commercial wide-field microscope was utilized for JT-SOFI super-resolution imaging of microtubules jointly labeled with QD525, QD625 and QD705 exhibiting inherent distinct blinking statistics. By controlling the sizes of quantum dots, one can synthesize different types of QDs with spectrally different fluorescence emissions. The QDs were further coated with a polymer shell that allows the materials to be conjugated to biological molecules and to retain their optical properties. Therefore, the resulting overall sizes of streptavidin Conjugates-polymer coated QDs are among the same range. All the three color QDs streptavidin Conjugates we used are around 15-20 nm. Also, the streptavidin covalently attached on the QDs surface are typically 5–10 streptavidins/Qdot conjugate, which results in all the QDs streptavidin conjugates with identical and a high specific biological activity. After imaging the joint-tagged microtubules, we have performed the analysis through averaging the total intensity divided by the intensity of the single QD to calculate the number of QDs in each channel, as shown in Fig. S5. The labeling density can also be reflected from each individual channel, which has been supplied in Supporting Information.

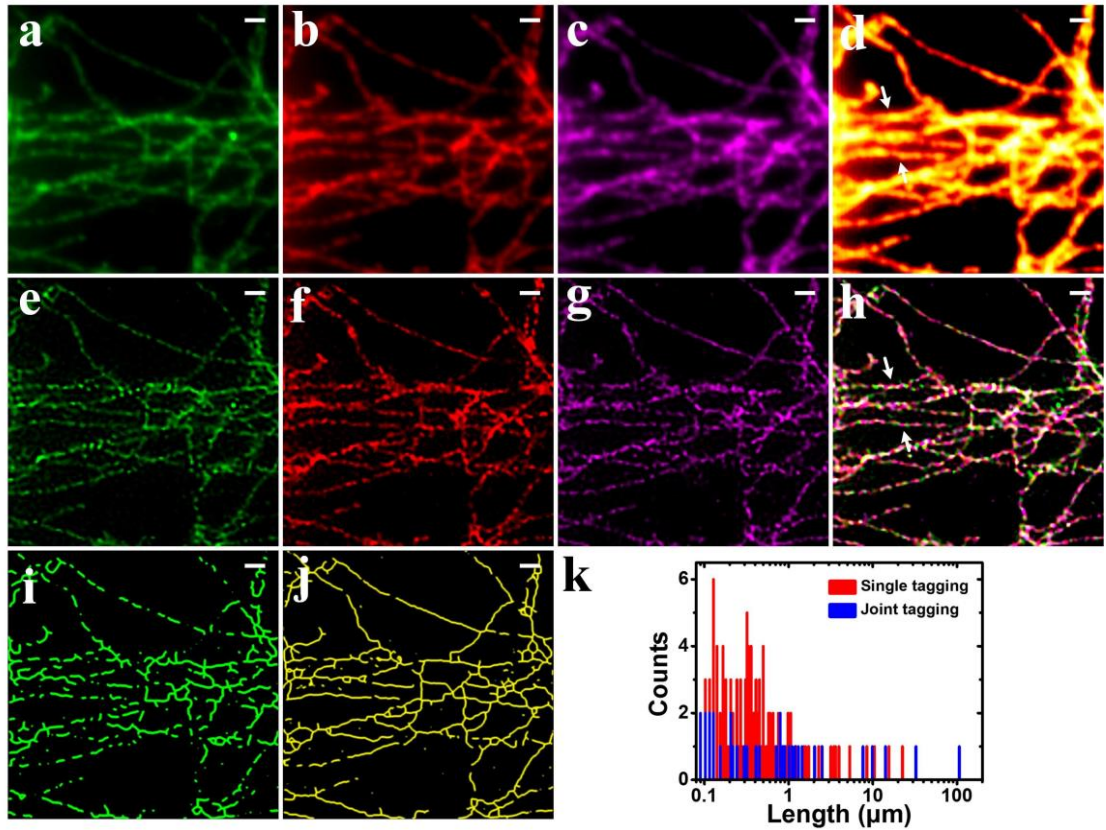

**Figure 3 Comparison of the imaging results of SOFI and JT-SOFI.** (a)-(c) Average images of microtubule networks labeled by QD525, QD625 and QD705, respectively. (d) Merged image of (a), (b) and (c). (e)-(g) 4th order SOFI images of microtubule networks labeled by QD525, QD625 and QD705, respectively. (h) Merged image of (e), (f) and (g). (i), (j) Skeletonized images of (e) and (h), in which the long continuous microtubules are predominant in (h). (k) Histograms of the length of the continuous lines in (i) and (j) are plotted. *Scale bars: 1 μm*.

In order to reduce the acquisition time, thus increasing the imaging speed, we only used 100 frames for high-order SOFI reconstruction in each channel. As the labeling density for each channel is relatively low, after de-noising, we found that 100 frames were sufficient for reconstructing a super-resolution image with high imaging fidelity through joint-tagging SOFI.

Fig. 3d was obtained by averaging Figs. 3a-c. Due to the diffraction limit, fine structures at the sub-diffraction level were blurred. Subsequently, SOFI algorithm was applied to the 3 sequences of images. The SOFI-processed images (Figs. 3e-g) show spatial resolution improvements over the average counterparts. However, the microtubules exhibit discontinuities and heterogeneities after 4th order SOFI processing. Afterwards, the 3 SOFI images were combined (Fig. 3h) and the discontinuities diminished significantly. Eventually, we transformed the composite image into single color image shown in Fig. 4b. As can be seen, the image very well retains the continuities of the microtubule networks without sacrificing the spatial resolution enhancement. The line profiles shown in Fig. 4c were plotted according to the arrows in Fig. 3d and 3h. The closely spaced microtubules whose fine structures are indistinguishable in average image can be well discerned by JT-SOFI. Fig. 4d shows that the spatial resolution can be improved from 210 nm to 85 nm with considerably reduced discontinuities and heterogeneities by employing joint tagging 4th order SOFI processing.

To compare the discontinuity qualitatively, we skeletonized the obtained microtubule networks image with a plugin in Fiji[28, 29], as shown in Figs. 3i and j. Apparently, the microtubule skeletons in Fig. 3j exhibit enhanced continuities and integrity over the counterpart in Fig. 3i. The lengths of continuous microtubule skeletons are further analyzed quantitatively, as shown in the histograms in Fig. 3k. This was done by a custom-written algorithm using Matlab. Obviously, Fig. 3i contains more numbers of skeletons with shorter lengths ranging from 0.1 to 10 μm. Whereas, Fig. 3j possesses more numbers of skeletons with the lengths stretching to tens of microns, even above 100 μm. The statistical analysis of skeletonized results revealed the robustness and superiority of JT-SOFI for achieving satisfying super-resolution imaging whilst preserving the integrity and continuities of the nanoscale structures under investigation.

We also performed live cell simulation, and experimentally demonstrated the capability of JT-SOFI for live cell imaging, revealing the dynamics of lipid rafts tagged with dual-color QDs in living cells (Figure S15).

In this work, the multi-color probes are jointly tagged on the same sub-cellular organelle. With more spectra employed, one may be able to perform additional labeling of other organelles, which enables the visualization of the interaction with super-resolution. Taking advantage of the narrow spectral width of QDs, the mixing of up to 5 QDs have been demonstrated previously[27]. Further, if one of them can be labeled at single molecule level (by adding an additional channel), then our method is so far the fastest and most faithful, through the combination of SOFI and single molecule localization. Also, it should be noted that, although here we employed QDs as the fluorescent label, other dyes which exhibit blinking can also be employed, such as photo-convertible organic dyes[30], plasmonic metal nanoparticles[31], etc. Moreover, the joint tagging is not restricted to spectral separation only. Other forms of differentiation can also be used toward a higher multiplexing degree, for example, the dipole orientation of the fluorescent dyes[32], or through lifetime differences[33]. It can also be used in conjugation with other optical imaging modalities such as total internal reflection fluorescence (TIRF) or confocal microscopy.

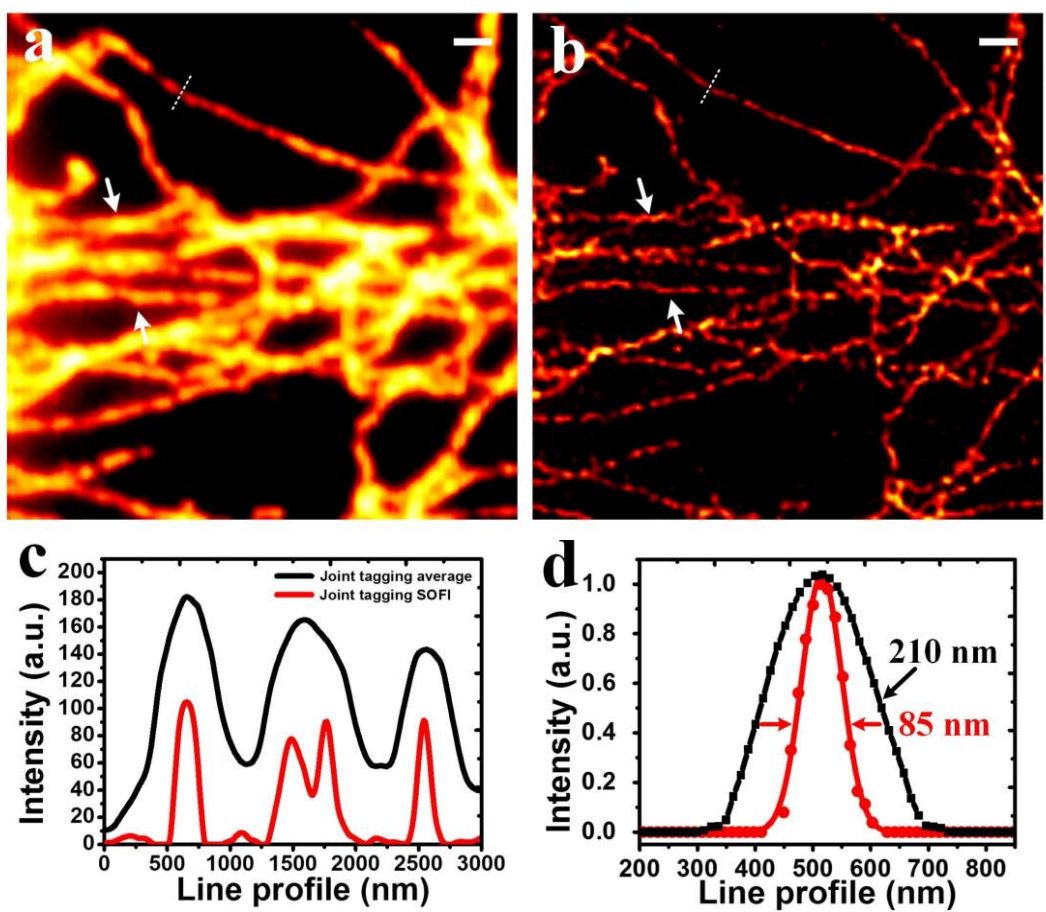

**Figure 4 Comparison of the wide-field and JT-SOFI.** (a) Wide-field merged average image from all three spectral channels. (b) 4th order JT-SOFI image. (c)

Cross-sections indicated by white arrows in (a) and (b). (d) Cross-sections indicated by white dashed lines in (a) and (b). *Scale bar: 1 µm*.

**Discussion**

In conclusion, our quantum dots joint tagging scheme has successfully expanded the application scope of SOFI algorithm in the labeling density regime. Based on standard wide-field microscope, we put forward joint-tagging SOFI to obtain fast nanometric super-resolution fluorescence imaging with well suppressed distortions and enhanced image fidelity. The problematical downfall of visibility in SOFI images under ultra-high labeling densities can be satisfactorily rescued. Since the fluorescence emissions of multiple types of QDs are well separated and can be captured simultaneously, this joint tagging scheme also enables fast SOFI super-resolution imaging. The generations of heterogeneities and discontinuities induced by high-order SOFI processing have been effectively overcome, thereby surmounting the mathematical barrier bestowed by high-order SOFI algorithm. As JT-SOFI can be applied in a variety of blinking/switching dyes, we anticipate a wide application of this method.

The advantage of JT-SOFI over SOFI lies on that JT-SOFI can preserve more structural information over conventional SOFI, which works only with limited labeling density (12 µm$^{-1}$). It should be noted that, SOFI has already surpassed the labeling density of single-molecule localization-based techniques (PALM/STORM) [17], which typically work below the labeling density of 8 µm$^{-1}$. By applying 3 channels, JT-SOFI works well under the labeling density of 24 µm$^{-1}$. Therefore, the overall labeling density for JT-SOFI is over 3-fold higher than that of PALM/STORM. As a consequence, JT-SOFI is more suitable for *in vivo* cellular imaging, due to its high fidelity and fast imaging speed. In this work, we have used commercial available QDs for fluorescent labeling. It should be noted that, the statistics of the blinking/fluctuation nature of the QDs can largely determine the data acquisition speed for SOFI. Previously, through the modification of the fluctuation of the QDs, Watanabe et al. has reported real time SOFI microscopy [34]. This method can be employed in conjugation with our JT-SOFI, to improve both the temporal resolution and the fidelity of the super-resolution imaging. The disadvantage of SOFI and JT-SOFI is on the limited resolution improvement as artifacts become dominant with increased cumulant orders: PALM/STORM can always decrease the number of the effective emitter to reach single molecule level through excessive photobleaching, and then use very long acquisition time for image reconstruction (2-12 hours reported in Ref. [35]), with a typical resolution of 20 nm. As a comparison, SOFI solely relies on the uncontrolled blinking nature of the QDs, and the resolution is typically 50-100 nm[15]. Recently, with the application of compressed sensing, the reconstruction of STORM image with 100-500 frames has also been reported, with a resolution of 60 nm[36].

Another unique advantage of JT-SOFI lies on its low requirement on frame number, thus provides better temporal resolution. Although there is no minimum frame number requirement for SOFI processing, practically 1000-2000 frames are often used in order to reconstruct a super-resolution image with acceptable imaging fidelity for high-order SOFI process [15, 17, 30]. High frame numbers imply a long data acquisition time, thus confines fast imaging applications for live cell study. There are two constraints for the frame numbers of SOFI processing: (1) the labeling density. In the meantime of parallel detection in JT-SOFI, for each channel the labeling density can be also decreased n-folds, making each channel sparser. The discrete spatial distribution can decrease the requirement of frame numbers for SOFI reconstruction. We can see from Fig. 2 that, the JT-SOFI with three color low labeling density (3x8 µm$^{-1}$, Fig.2o) has much better result than the high labeling density counterparts (24 µm$^{-1}$, Fig. 2g). As shown in Fig. S9 in the Supplementary Information, increasing the frame number can significantly improve the connectivity, therefore 1000 frames were previously used in SOFI. The 3-color JT-SOFI obtained by 100 frames/channel (Fig. 2o) already exceeds the performance of 1000 frames conventional SOFI (Fig. S9j). (2) the blinking statistics. With enhanced blinking, Watanabe et al. have demonstrated super-resolution imaging with only 10 frames with modified SOFI variance imaging[34]. It should be noted that, the blinking statistics is related to not only the emitters' blinking ratio, but also the pixel size, as within the pixel size the signal intensity is averaged over one frame. Consequently, it is beneficial to use smaller pixel size to probe the rapid blinking of each QD or other fluctuating small molecules. Although it is suggested that for SOFI the pixel size of 150-170 nm is sufficient for high order SOFI [37], it is theoretically derived omitting the constraint of limited frame numbers. Recent reports of SOFI with much less frames are all based on the application of reduced pixel size ranging between 50-65 nm [34, 38]. For example, by using effective pixel size of 65 nm, Cho et al. have successfully obtained SOFI image with 300 frames, with organic dye in which the fluctuation is much less than that of QDs[38]. Here we also used the small pixel size (65 nm) to reduce the frame number yet producing reliable SOFI image.

Furthermore, the joint tagging multiplexing demonstrated in this work is not limited to SOFI, but rather, it can be applied in all sorts of stochastic localization-based microscopy [39, 40], in which the labeling density and imaging speed is always a conflict. Moreover, although here we have only demonstrated the application of multi-color QDs in JT-SOFI, the core concept is to separate the emitters first in another dimension and then recombine the results, such as fluorescent spectrum[27], dipole orientation [32], or lifetime [33]. This is predicted by Prof. Eric Betzig in 1995 [41], in which the optical properties of the emitters can be used to separate them to different dimensions toward single-molecule localization. JT-SOFI, however, can employ both the dimension and the statistics of fluctuation to yield high spatial and temporal resolution simultaneously.

**Methods**

The image collections of the microtubule networks was performed on a wide-field fluorescence microscope (DeltaVision, Applied Precision) with an oil objective (Olympus, 100x, Numerical aperture: 1.4). A laser with the wavelength of 405 nm was employed for simultaneously exciting the fluorescence emissions of QD525, QD625 and QD705 (power density: 120 W/cm$^2$). Three sets of filters were applied for separating the fluorescence emissions from QD525, QD625 and QD705, respectively. The fluorescence signals were acquired using cooled charge-coupled device (CCD) camera (CoolSNAP HQ2). The effective pixel size on the sample is 65 nm. The exposure time of each frame was 45 ms, and 100 frames were captured for each channel. During our experiment, no fiducial bead is needed for drift correction. After data acquisition, we implemented sub-pixel drift correction by employing the image registration algorithm based on nonlinear optimization and discrete Fourier transforms[42]. During the experiment shown in Figs. 3 and 4, the time-lapse images from the three channels were collected by sequentially switching three sets of dichroic filters due to the lack of three CCDs for simultaneous collection. 100 frames were captured for the green, red and magenta channel, respectively.

JT-SOFI requires the separated spectral detection of each channel. It can be done simultaneously with the employment of multiple CCD cameras and corresponding dichroic filters. It can also be done by splitting the detector into several regions, with each region corresponds to one spectral channel, at a sacrifice of the field of view [43], as demonstrated in the live cell experiment in this paper. It can also be done sequentially by switching the dichroic filters, and then detect the images, as demonstrated in the fixed cell experiment in this paper. An m-fold data acquisition time is needed in this case, comparing with the simultaneous detection.

In SOFI, the spatio-temporal cross-cumulant between neighboring pixels are calculated [37]. By calculating the cross-cumulants, an approximation of the underlying Point Spread Function (PSF) can be well estimated. This enables the straightforward processing of SOFI images, eliminating shot-noise and camera read-out noise. Furthermore, by using maximum likelihood estimation (MLE) following with reconvolution, the resolution can be linearly improved over cumulant order[44]. Additionally, by estimating the blinking statistics of the fluorophores using balanced SOFI (bSOFI)[44], the nonlinear response to brightness and blinking heterogeneities can be effectively eliminated. This significantly enhances the visual perception of high-order SOFI images. In this work, SOFI processing is performed with the open-source software bSOFI[44].

However, although such mathematical efforts have been exerted, when high-order SOFI algorithm is applied to biological applications with ultra-high labeling densities directly, mathematical artifacts always degrade the image quality of the obtained super-resolution images. Therefore, we propose a method to overcome the problem by multi-color joint tagging. By introducing this method, for one spectral channel, we indirectly reduce the labeling density for high-order SOFI algorithm to be fully performed with minimized induced artifacts and discontinuities. After the powerful combination of the multi-channel results, we are able to produce super-resolution images with well-preserved integrity and continuities in the ultra-high labeling density regime.

We use local optimum thresholding for image binarization of the microtubules[45]. A local version of Otsu's global threshold clustering was implemented. The algorithm searches for the threshold that minimizes the intra-class variance, defined as a weighted sum of variances of the two classes. The local set is a circular ROI and the central pixel is tested against the Otsu threshold found for that region. After that, a plugin of Fiji called Skeletonize3D was used to extract the skeleton of the microtubules[28, 29]. Skeletonization is a morphological image processing method which can provide a simple and compact representation of a structure preserving many of the topological and geometrical characteristics of the original structure[46]. The analyses of the numbers and lengths of the continuous microtubule skeletons were done with morphological binary image object label. It is processed by first using Matlab Image Processing Toolbox's morphological image processing tool, bwlabel, to obtain the 8-connective regions (the neighborhood connective pixels are recognized as one object). After labeling the binary image of the microtubule, the numbers of the pixels in each connective skeleton regions are calculated as the length of an isolated skeleton. The code of the custom-written Matlab program is listed in the Supporting Information.

**Immunostaining of microtubules in COS7 cells.** COS7 cells were seeded on glass cover slides for overnight growth. Before staining, cells were washed with 1X PBS buffer, and extracted for 1 min with 0.2% Triton X-100 in a pH 7 buffer consisting of 0.1 M PIPES, 1 mM ethylene glycol tetraacetic acid, and 1 mM magnesium chloride. Then cells were fixed in 4% paraformaldehyde (Electron Microscopy Sciences) and 0.1% glutaraldehyde (Electron Microscopy Sciences) in PBS for 10 min. Reduction was followed with 1mg/ml NaBH4 for 5 min and then wash with PBS. The cells were blocked and permeabilized with 5% bovine serum albumin (Jackson ImmunoResearch Laboratories) and 0.5% v/v Triton X-100 in PBS for 30 min. After blocking and permeabilization, anti-alpha tubulin primary antibody with biotin (ab74696, Abcam) was diluted to 10 µg/mL in blocking buffer and added to cells for 40 min. After washing with PBS, cells were stained for 60 min with 3 color QDs streptavidin conjugates (Invitrogen) with emission peak at 525, 625, 705nm, and each color QDs were diluted to 15nM in blocking buffer. Then cells were washed with PBS for 3 times, 5 minutes each. Finally a post-fixation was carried out in a mixture of 4% formaldehyde and 0.1% glutaraldehyde in PBS for 10 min. Then cells was washed and mounted with 50% glycerin.

 **Acknowledgments**

This work was supported by the National Instrumentation Program (2013YQ03065102), the "973" Major State Basic Research Development Program of China (2011CB809101, 2010CB933901), and the National Natural Science Foundation of China (61178076, 31327901, 61475010).

**Author contributions**

X.C., P.X. and Z.Z. conceived the project. Z.Z. and H.Z. carried out the simulation. N.H., C.S. and J.T. prepared the live cell specimens. Z.Z., X.C., and H.W. performed the imaging experiments and analyzed the data. Z.Z., X.C., and P. X. wrote the manuscript with input from all authors.

**Additional information.**

Competing financial interests: The authors declare no competing financial interests.

# Supplementary Information

## Fast Super-Resolution Imaging with Ultra-High Labeling Density Achieved by Joint Tagging Super-Resolution Optical Fluctuation Imaging

Zhiping Zeng[1], Xuanze Chen[1], Hening Wang[1], Ning Huang[2], Chunyan Shan[2], Hao Zhang[1], Junlin Teng[2] & Peng Xi[1]*

[1]Department of Biomedical Engineering, College of Engineering, Peking University, Beijing, China;

[2]College of Life Sciences, Peking University, Beijing, China

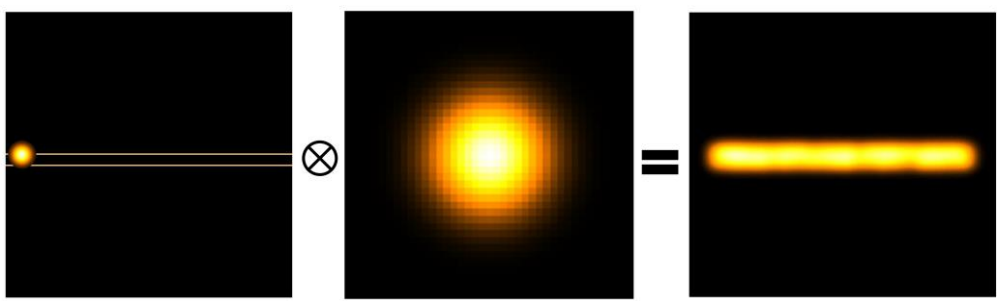

**FIGURE S1 Simulated two closely spaced parallel lines (within diffraction limit) convolved with a Gaussian point spread function, generating a blurred image in which two parallel lines cannot be resolved anymore. This simulation illustrates the image formation originating from a diffraction-limited optical system. Each emitting point on these two parallel lines becomes a light spot on the image plane, thereby hindering us from discerning the fine structures.**

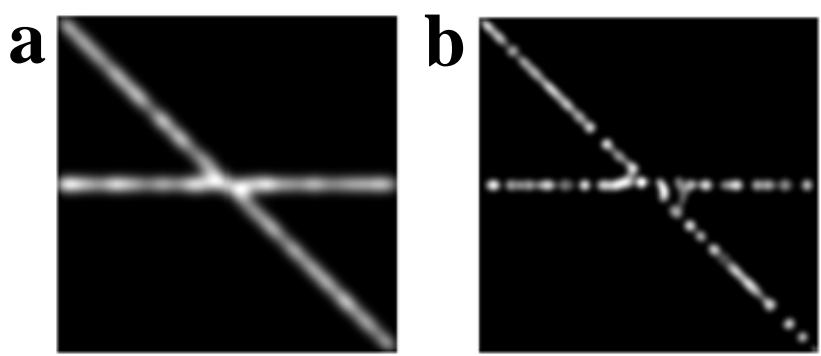

**FIGURE S2 Simulation of two cross lines with high labeling density (24 μm$^{-1}$) blinking emitters. And only 100 time slots are simulated. (a) The 2nd order SOFI presents less**

resolution improvement, but the two cross lines are continuous. (b) The 4th order SOFI achieves better resolution enhancement, meanwhile, discontinuities have also been undesirably generated. Image size: 8 μm × 8 μm. Pixel sizes are 40 nm for (a) and 20 nm for (b), respectively.

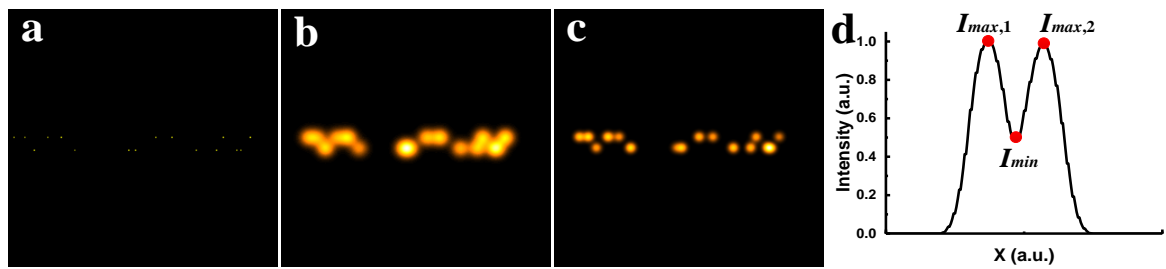

**FIGURE S3** Schematic illustration of the visibility defined in Eq. (1). (a) The target distribution. (b) The average image of the targets. (c) 3rd order SOFI image of the targets. (d) Line profile obtained by projecting the SOFI image along horizontal direction. $I_{max,1}$ and $I_{max,2}$ are two normalized peak values along the profile. $I_{min}$ is normalized minimum value along the profile.

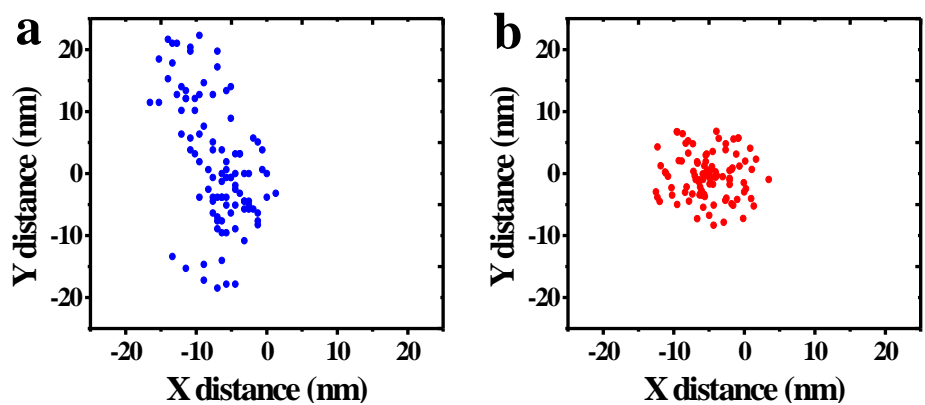

**FIGURE S4** Before (a) and after (b) system drift correction of 100 frames. The subpixel drift correction algorithm is able to correct the system drift based on the raw data[1]. No fiducial bead is needed for drift correction. During the experiment, the system drift was caused by various factors, e.g., instability of the stage, vibrations from peripheral equipment and machinery, other unobservable disturbances, etc. Before correction, the drift of the system is indicated by blue dots in (a). As can be seen, the maximum drift along Y direction is over 40 nm. After correction, the drift can be confined within a range of less than 15 nm, indicated by red dots in (b).

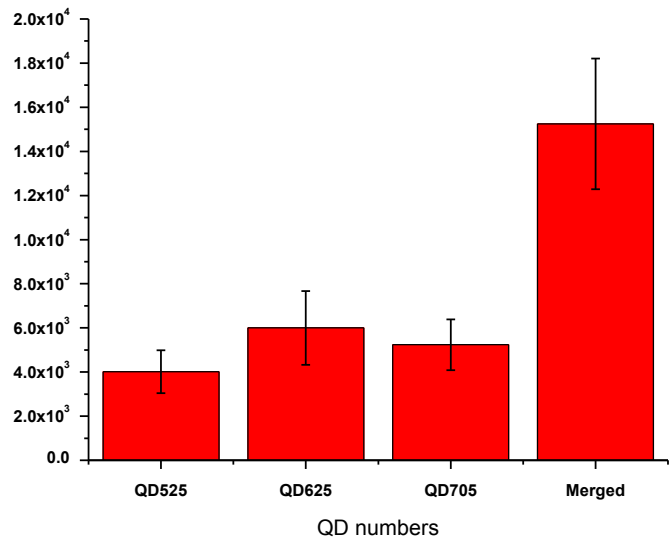

**FIGURE S5 Estimated numbers of quantum dots in Figure 3e, 3f, 3g and 3h, respectively.**

Since the estimated numbers of QDs in three channels range from roughly 4000 to 6000, and the estimated total lengths of the microtubules range from 136 μm to 149 μm. Thus, taking into account of the volumetric distribution of the QDs labelled to the microtubules, we estimated the density in each color channel is around 7.3 $\mu m^{-1}$ to 10.1 $\mu m^{-1}$.

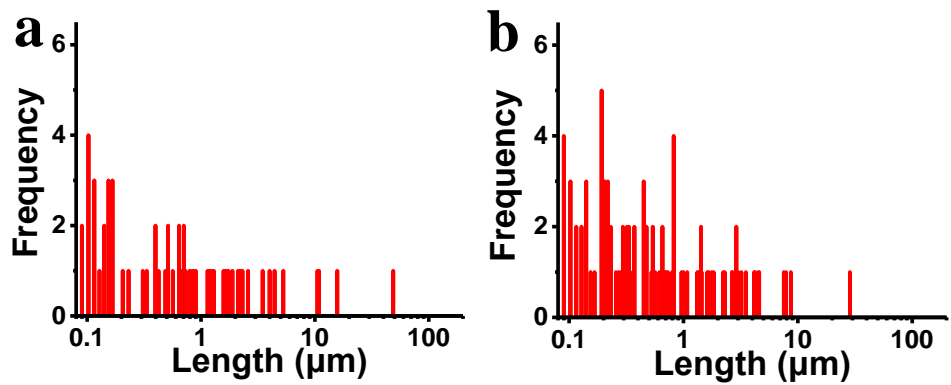

**FIGURE S6 (a) Histogram from the skeletonized image shown in Figure 3f. (b) Histogram from the skeletonized image shown in Figure 3g.**

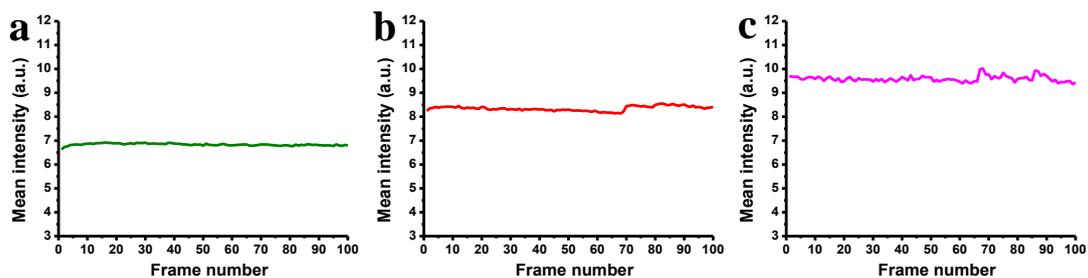

**FIGURE S7 Mean fluorescence intensity variations versus frame numbers for (a) QD525, (b) QD625 and (c) QD705. The photobleaching is negligible after capturing 100 frames for all three types of QDs. Due to the photostability of the QDs, photobleaching is not noticeable.**

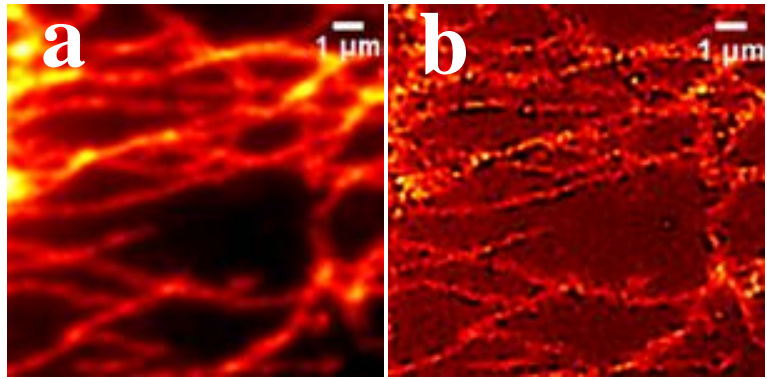

**FIGURE S8 Comparison of single color labeled microtubules in possible maximum density. (a) Averaged image. (b) Conventional 4th order SOFI image.**

Fig. S8 shows the microtubule networks under the labelling density of 24 $\mu m^{-1}$. Fig. S8a is averaged image, Fig. S8b is the conventional 4th order SOFI result. As can be seen, under such high labelling density, without the implementation of JT-SOFI, 4th order SOFI fails to present an acceptable high-resolution image. The generated image suffers from severe artifacts and discontinuities rather than high-quality super-resolution image that JT-SOFI has demonstrated.

**JT-SOFI can decrease the requirement for large frame numbers in SOFI processing, thereby improving spatial-temporal resolution simultaneously.**

Increasing the frame number can significantly improve the connectivity, as the more frame number, the chances of ON states are also increased linearly. To demonstrate the effect of high-order SOFI processing with different labeling densities and frames, we have simulated the parallel lines with distance of 200 nm, and labeling density of 8 μm$^{-1}$(upper) and 24 μm$^{-1}$(lower), respectively, as shown in Fig. S9. When the density is relatively low (8 μm$^{-1}$), 100 frames for reconstruction is able to present reliable super-resolution image. Whereas, when the density is high (24 μm$^{-1}$), images generated using less frame numbers (e.g., 100 and 200 frames) contain significant discontinuities. Only images reconstructed using sufficient frame numbers (e.g., 500 and 1000 frames) preserve the continuities of structure. As a consequence, the frame number required for SOFI reconstruction is significantly dependent on the labeling density. Low density labeling is capable of dramatically reducing the frame number, notably expediting the imaging speed for SOFI imaging.

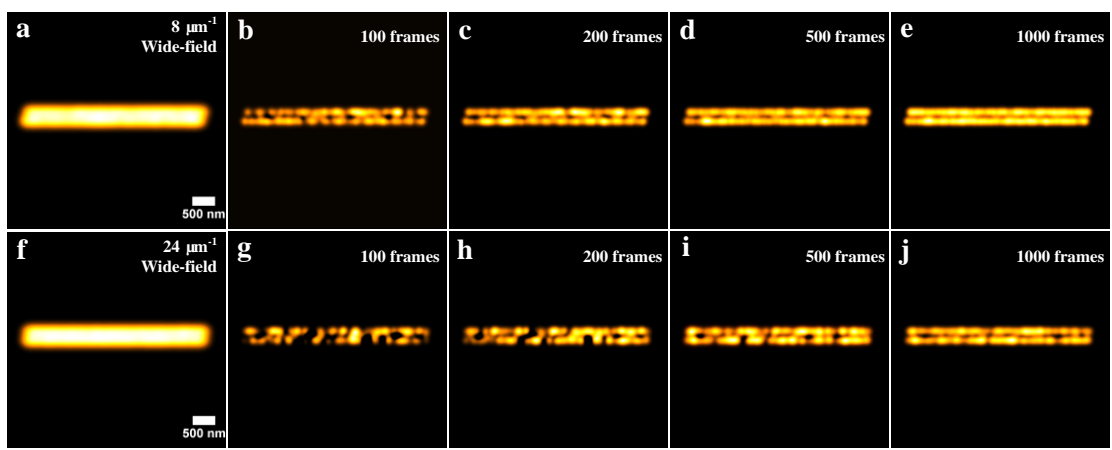

**FIGURE S9 (a) Average image of two closely spaced parallel lines distributed with blinking emitters with the labeling density of 8 μm$^{-1}$. (b-e) 3rd order SOFI images using 100, 200, 500 and 1000 frames for reconstruction, respectively, for labeling density of 8 μm$^{-1}$. (f) Average image with the labeling density of 24 μm$^{-1}$. (g-j) 3rd order SOFI images using 100, 200, 500 and 1000 frames for reconstruction, respectively, for labeling of 24 μm$^{-1}$.** *Scale bars: 500 nm.*

In an extreme case, we have also simulated the SOFI reconstruction with only 10 frames, in different labeling density situations, as shown in Fig. S10. The upper and lower panels are for labeling density of 12 μm$^{-1}$ and 3 colors with each 4 μm$^{-1}$, respectively. As the labeling density is low, the 10-frame averaged wide-field result for 12 μm$^{-1}$ (Fig. S10a) is better than that of 4

μm$^{-1}$ (Fig. S10d). However, the 3-channel JT-SOFI result (Fig.10f) demonstrates even better connectivity than the 12 μm$^{-1}$, 100 frames result (Fig. S10c).

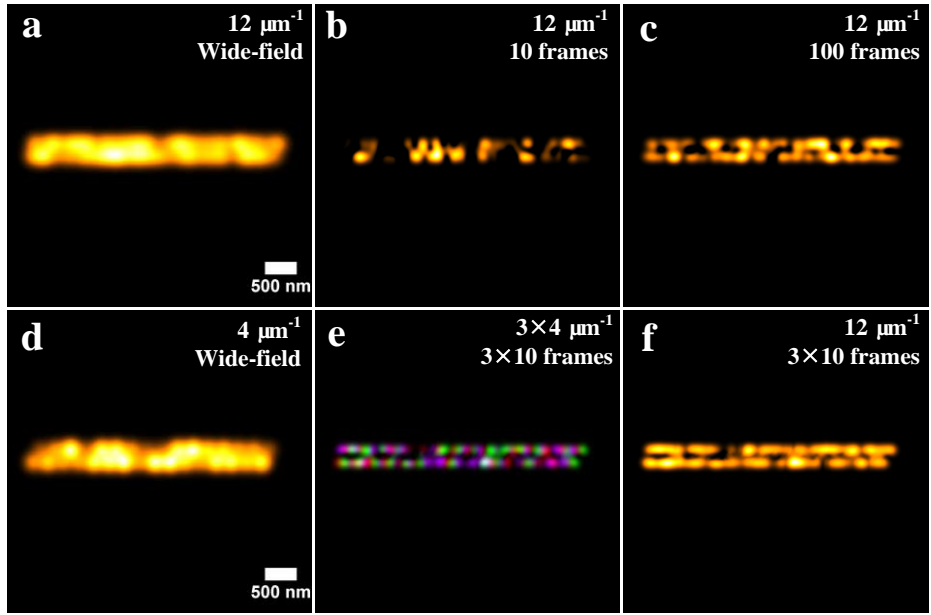

**FIGURE S10 (a) Average wide-field image of two closely spaced parallel lines distributed with blinking emitters with the labeling density of 12 μm$^{-1}$. (b) 3rd order SOFI image using 10 frame for reconstruction with the labeling density of 12 μm$^{-1}$. (c) 3rd order SOFI image using 100 frame for reconstruction with the labeling density of 12 μm$^{-1}$. (d) Average wide-field image with the labeling density of 4 μm$^{-1}$. (e) 3-color joint tagging with the labeling density of 4 μm$^{-1}$ for single channel and 10 frames were used for SOFI reconstruction for each channel. (f) 3rd order SOFI image after summing up 3 channels with the overall labeling density of 12 μm$^{-1}$.** *Scale bars: 500 nm.*

In Fig. S11, we simulated images with different pixel sizes, at labeling density of 8 μm$^{-1}$. As smaller pixel size can detect more fluctuation, while larger pixel size averages the fluctuation over space, it can enable SOFI imaging using significantly reduced frame numbers (200 frames) with better image quality (Fig. S11c) over the counterparts with larger pixel size, and 1000 frames (Fig. S11j).

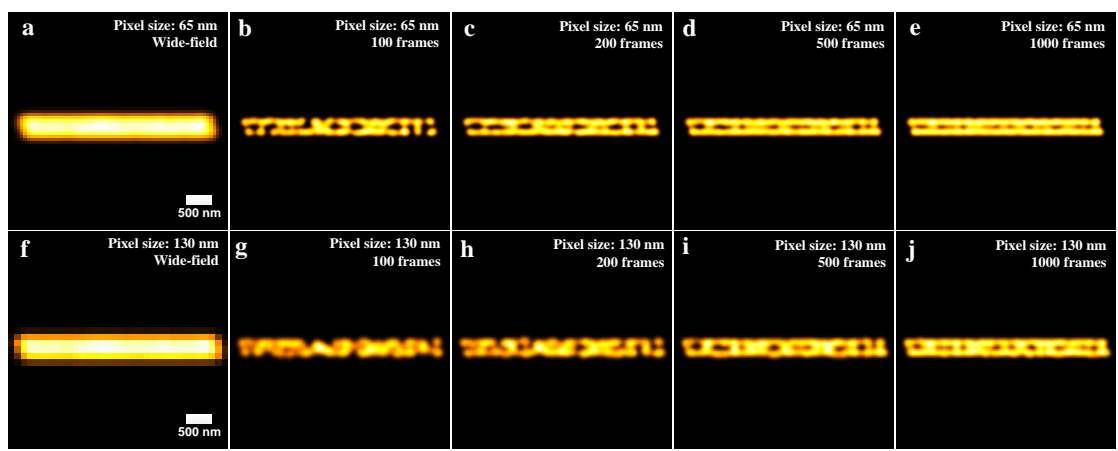

**FIGURE S11** (a) Average image of two closely spaced parallel lines distributed with blinking emitters with the pixel size of 65 nm. (b-e) 3rd order SOFI images using 100, 200, 500 and 1000 frames for reconstruction, respectively, for pixel size of 65 nm. (f) Average image with the pixel size of 130 nm. (g-j) 3rd order SOFI images using 100, 200, 500 and 1000 frames for reconstruction, respectively, for pixel size of 130 nm. *Scale bars: 500 nm.*

**Simulation of real-time imaging of living cells.**

In this simulation, a dynamic microtubule network was simulated for conventional SOFI and JT-SOFI imaging. The overall labeling densities in conventional SOFI and JT-SOFI were set identical to 24 $\mu m^{-1}$. 300 frames were collected for conventional SOFI imaging. For JT-SOFI imaging, 100 frames were collected for each channel, which corresponds to 300 frames for three channels. The dynamic motion was simulated through rotating the microtubule network by 0.01 degree during each frame, resulting in 1 degree after JT-SOFI collection and 3 degrees after conventional SOFI collection, the rotation orientation is indicated by the white clockwise arrow in Fig. S13a. Fig. S13a shows the averaged result by averaging 300 frames collected from three channels by joint tagging. Fig. S13b presents the result from conventional SOFI processing by collecting 300 frames using one single channel. Apparently, owing to the dynamic motion of the microtubule network during acquisition, conventional SOFI gives a severely blurred image, as indicated in magnified images in Figs. S13e and S13h. Nonetheless, JT-SOFI image in Fig. S13c presents a much better result. As can be seen, in magnified images indicated in Figs. S13f and S13i, the blurring is almost negligible compared to Figs. S13e and S13h. Furthermore, the resolution enhancement obtained from JT-SOFI is more significant than that from conventional SOFI. From the cross-sections shown in Fig. S13j, JT-SOFI is able to discriminate the sub-diffraction interleaving structure indicated by the white dashed lines in Figs. S13d, S13e and S13f. However, due to the diffraction limit and limited resolution, the averaged image cannot distinguish this structure. Moreover, conventional SOFI also fails to resolve this structure. This can be ascribed to the severe blurring and reduced resolution improvement caused by the insufficient imaging speed and incompetent high labeling density implementation when conventional SOFI is applied to living cell imaging.

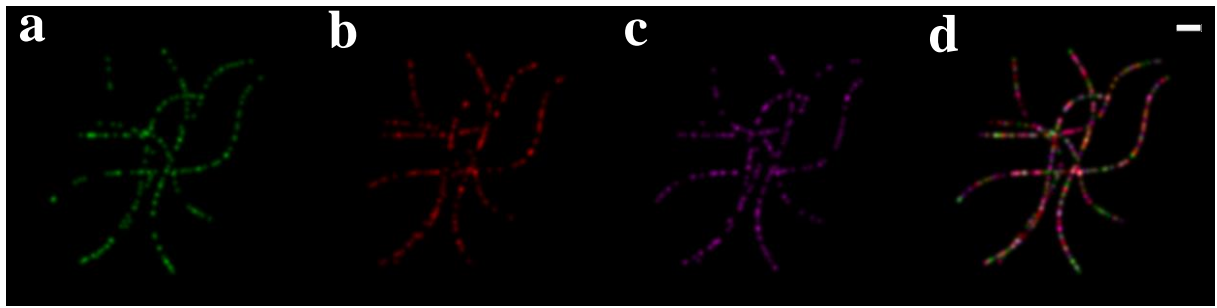

**Figure S12 Single frames of the raw images in living cell simulation. (a) The image from the green channel. (b) The image from the red channel. (c) The image from the magenta channel. (d) Merged image of the three channels.** *Scalebar: 1 μm.*

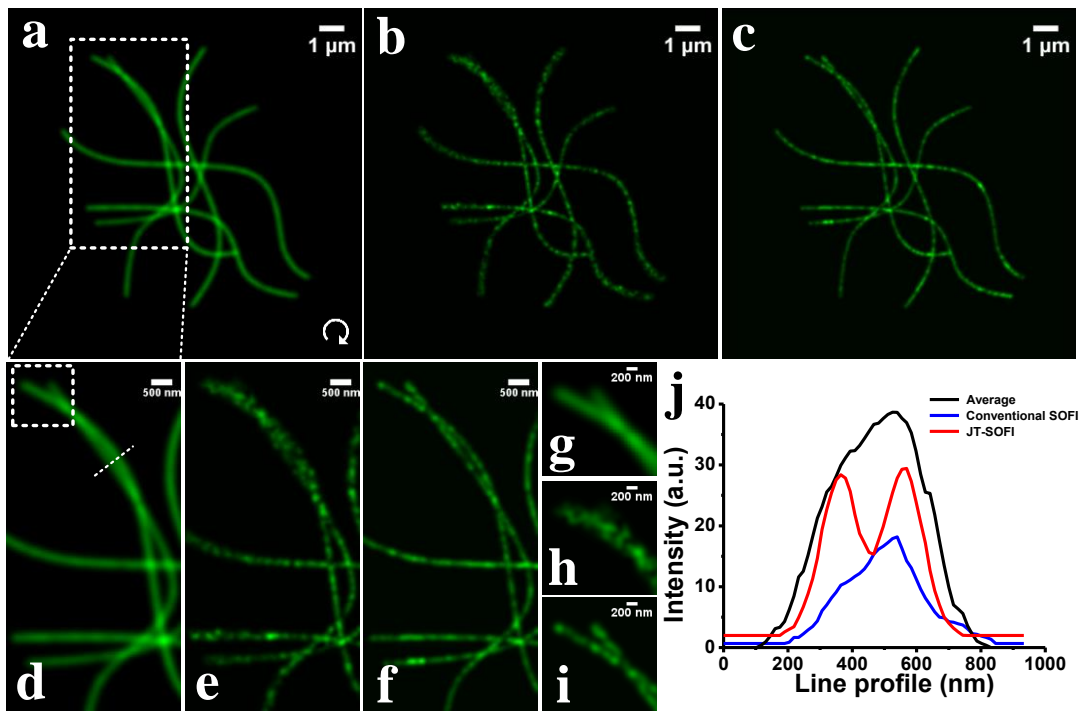

**Figure S13** Simulation of real-time imaging of living cells. (a) Average image. (b) Conventional SOFI image. (c) JT-SOFI image. (d-f) Magnified regions from the white boxed regions in (a-c). (g-i) Small regions outlined by the white boxes in (d-f). (j) Cross-sections indicated by the white dashed lines in (d-f).

**JT-SOFI imaging reveals the dynamics of lipid rafts tagged with dual-color QDs in living cell.**

To demonstrate the potential application of JT-SOFI to living cell imaging, we observed lipid rafts tagged with dual-color QDs in living COS7 cells, and single-particle tracking SOFI (sptSOFI) has been implemented[2]. In this experiment, QD525 and QD655 were jointly tagged to the lipid rafts. In order to simultaneously collect the fluorescence signals from QD525 and QD655 with one single CCD camera, we employed an image splitter to divide the CCD into two independent regions for simultaneous detection of the two channels. The dual-color QDs were simultaneously excited by 488 nm laser. The exposure time of the camera was set at 18 ms. The time-lapse sequences were obtained by continuously capturing the two channel images for hundreds of frames. And each 100 frames were extracted for reconstructing a SOFI image at a particular time point. Figs. S15a-b are the wide-field image and the corresponding 3rd SOFI image from 100 frames in the QD525 channel. Figs. S15c-d are the wide-field image and the corresponding SOFI image from 100 frames in the QD655 channel. As can be seen, the out-of-focus blur can be well rejected and the spatial resolution was enhanced after SOFI processing. Further, we observed the dynamic motion of single QD-tagged lipid raft indicated by the dotted white circles in Figs. S15a-b. Fig. S15e presents five average images of the labeled lipid raft during a time course from 0 to 7.2 seconds. From a merged image of the 1st and 5th images in Fig. S15e, it is difficult to elaborate the dynamic movement of the lipid raft. However, from the SOFI counterparts shown in Fig. S15f, the dynamic movement of the labeled lipid raft can be evidently observed. We estimated that the lipid raft underwent a movement of ~100 nm along horizontal axis during 7.2 s, which was unavailable from the average image sequences. Furthermore, with the implementation of JT-SOFI, the labeling density can be increased by m-folds, consequently the lipid rafts that can be tracked are increased as well. Figs. S15g and S15h present the average and SOFI results in the QD655 channel. Due to the low SNR of this channel, 200 images were used for $2^{nd}$ order SOFI processing.

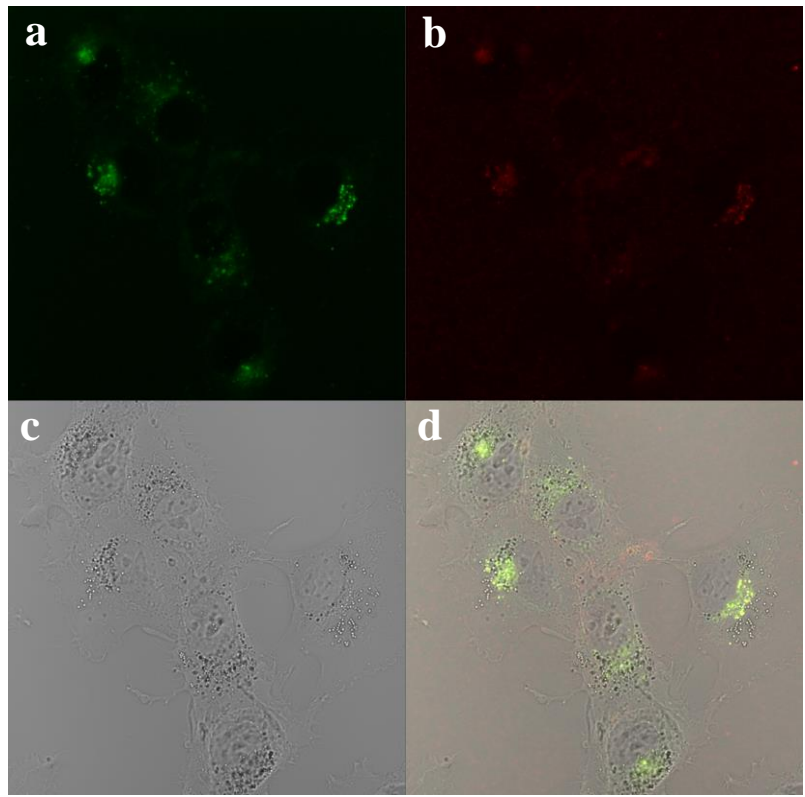

**Figure S14 Confocal and bright-field images of the lipid rafts in living COS7 cells. (a) Lipid rafts labeled by QD525. (b) Lipid rafts labeled by QD655. (c) Bright-field image of living COS7 cells. (d) Merged image of (a), (b) and (c).**

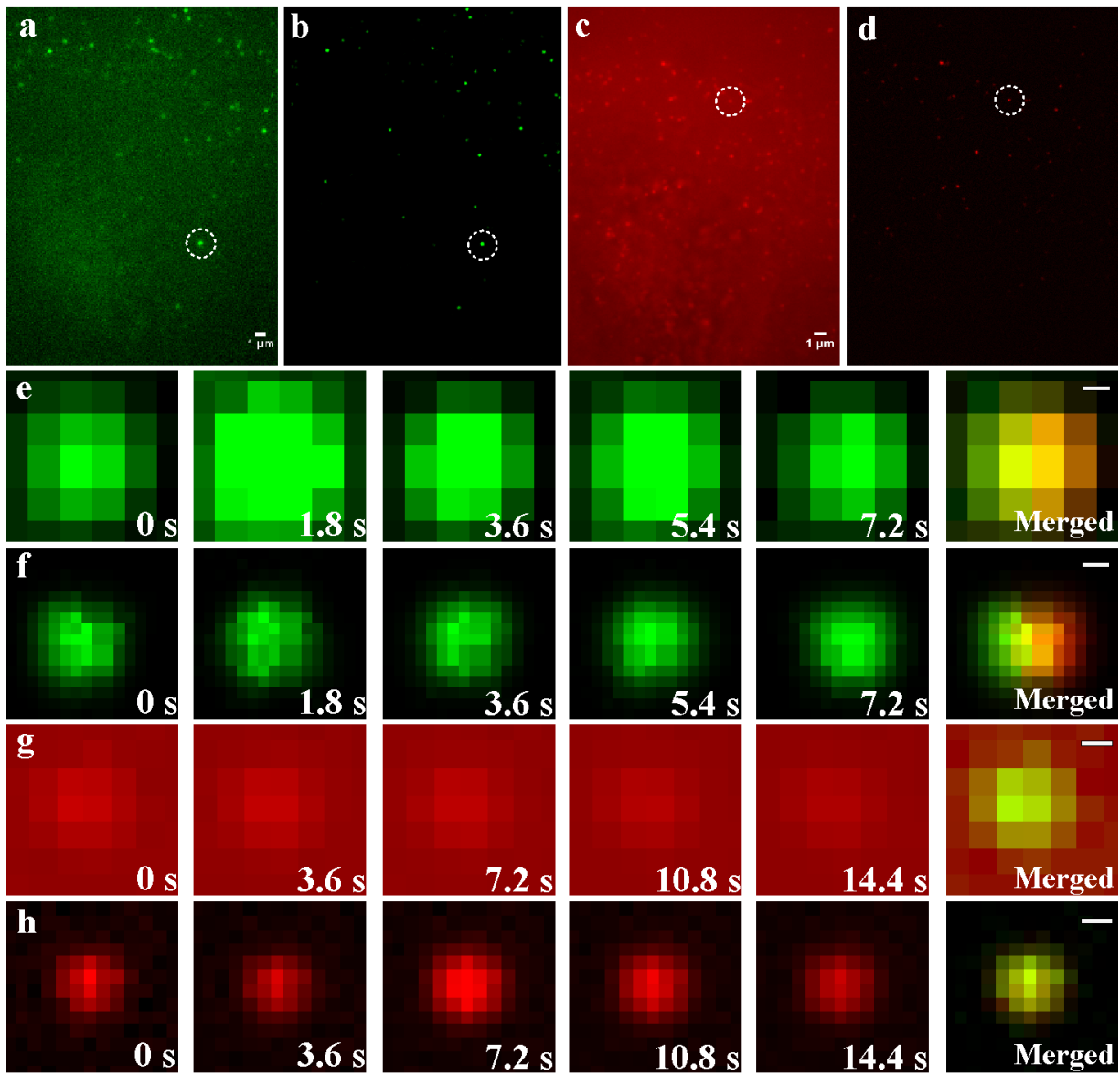

**Figure S15 Wide-field and JT-SOFI images of lipid rafts tagged with dual-color QDs in living COS7 cell.** (a) Wide-field image of lipid rafts labeled by QD525. (b) 3rd SOFI image of lipid rafts labeled by QD525. (c) Wide-field image of lipid rafts labeled by QD655. (d) 2nd SOFI image of lipid rafts labeled by QD655. *Scale bars in (a,c): 1 μm.* (e,f) Magnified regions indicated by the dotted white circles in (a,b) showing the time-lapse average and SOFI images of a lipid raft. The 6th images in (e,f) are the merged images of the corresponding 1st and 5th images. The color of the 5th images were reassigned as 'red' to distinguish from the 1st images when merged. (g,h) Magnified regions indicated by the dotted white circles in (c,d) showing the time-lapse average and SOFI images of a lipid raft. The 6th images in (g,h) are the merged images of the corresponding 1st and 5th images. The color of the 1st images were reassigned as 'green' to distinguish from the 5th images when merged. *Scale bars in (e-h): 100 nm.*

**Image acquisition.**

The image acquisition of the lipid rafts in live COS7 cells (African green monkey fibroblast) was carried out on a wide-field microscope (Olympus IX71) with an oil objective (100x, Numerical aperture: 1.49). A laser with the wavelength of 488 nm was used for simultaneously exciting the fluorescence of both QD525 and QD655. The images were captured by an EMCCD (Andor iXon Ultra), which was split into two regions by an image splitter (OptoSplit) for simultaneous detection of dual-color fluorescence signals. The exposure time of the EMCCD was set at 18 ms.

**Live cell culture and labeling protocol of the lipid rafts by quantum dots.**

COS7 cells were seeded on a glass-cover-slip-bottomed cell culture dish (NEST) for overnight growth at 37 ℃ under 5% $CO_2$. For labeling the lipid rafts, COS7 cells were first incubated with 1 μg/ml cholera toxin subunit B (recombinant), biotin-XX conjugate (Invitrogen) for 30 min before a 10 min 0.5 nM Qdot® 525 and 655 streptavidin conjugate (Invitrogen) incubation, then washed the cells by PBS three times and added the fresh culture medium. After culturing for 0.5-1 h, the endocytosis of the lipid rafts labeled by quantum dots was detected.

**TABLE S1 A list of the mean microtubule track lengths in different channels.**

|              | Mean track length (μm) | Maximum track length (μm) |
|--------------|------------------------|---------------------------|
| QD525        | 0.736                  | 22.528                    |
| QD625        | 1.128                  | 48.750                    |
| QD705        | 0.842                  | 28.544                    |
| Multi-color  | 2.283                  | 106.698                   |

**TABLE S2 Comparison of the conventional diffraction limited and super-resolution microscopy techniques.**

|                      | Wide-field  | Confocal    | JT-SOFI           | STED              | SIM               | PALM/STORM         |
|----------------------|-------------|-------------|-------------------|-------------------|-------------------|--------------------|
| **Imaging speed**    | Video-rate  | Video-rate  | seconds per frame | seconds per frame | seconds per frame | minutes per frame  |
| **Resolution (nm)**  | ~ 250       | ~ 200       | ~ 80              | < 50              | ~ 120             | ~ 20               |
| **Cost**             | Low         | Moderate    | Low               | High              | Moderate          | Moderate           |

In Fig. 2p, we simulated the visibility of SOFI reconstructed images versus labeling densities. Here, the visibility is defined as

$$v = \frac{I_{max,1} - I_{min}}{2(I_{max,1} + I_{min})} + \frac{I_{min} - I_{max,2}}{2(I_{min} + I_{max,2})}. \quad (1)$$

where $I_{max,1}$, $I_{max,2}$ and $I_{min}$ are the local maximum and minimum of the corresponding ROI.

# The custom-written Matlab program code for analyzing the numbers and lengths of the continuous microtubule skeletons.

```matlab
%% Calculate the numbers and lengths of the input binary images
function [y,leng] = calNum( img)
% img is a binary image matrix;
% y is the number of lines except background
[L, num] = bwlabel( img, 8);
L1 = L.* img;
temp = num;
for kk = 1:num
    index = find( L1 == kk);
    if( length(index) == 0)
        temp = temp - 1;
    end
end
y = temp;
for kk = 1:num
    index = find( L1 == kk);
    leng(kk)=length(index);
end

%% Plot histograms for the skeletonized images
close all
clear all
clc
A1 = imread('525_skeleton.tif');
B1 = imread('multi_color.tif');
A2 = im2bw(double(A1));
B2 = im2bw(double(B1));
[y1,leng1] = calNum( A2);
[y2,leng2] = calNum( B2);
figure
x1 = 1:1:4000;
```

```
[N1,X1]=hist(leng1,x1);
axis([-10 4000 0 75])
h = findobj(gca,'Type','patch');
set(h,'FaceColor','r','EdgeColor','w')
figure
x2 = 1:1:4000;
[N2,X2]=hist(leng2,x2);
axis([-100 4000 0 70])
h2 = findobj(gca,'Type','patch');
set(h2,'FaceColor','r','EdgeColor','w')
```

**SUPPORTING REFERENCES**